\documentclass[figures]{epl}
\usepackage{graphicx}

\title{Identification of Coulomb blockade and macroscopic quantum tunneling 
by noise}
\shorttitle{Identification of CB and MQT by noise}
\author{H. Grabert\inst{1}\thanks{E-mail: 
        \email{grabert@physik.uni-freiburg.de}}
\and G.-L. Ingold\inst{2}\thanks{E-mail: 
        \email{gert.ingold@physik.uni-augsburg.de}}}
\shortauthor{H. Grabert \etal}
\institute{
  \inst{1} Fakult{\"a}t f{\"u}r Physik, Albert-Ludwigs-Universit{\"a}t,
           Hermann-Herder-Stra{\ss}e 3, D-79104 Freiburg, Germany\\
  \inst{2} Institut f{\"u}r Physik, Universit{\"a}t Augsburg,
           Universit{\"a}tsstra{\ss}e 1,\\ D-86135 Augsburg, Germany
}

\begin{document}

\maketitle

\begin{abstract}
The effects of Macroscopic Quantum Tunneling (MQT) and Coulomb Blockade (CB)
in Josephson junctions are of considerable significance both for the
manifestations of quantum mechanics on the macroscopic scale and
potential technological applications. These two complementary effects are
shown to be clearly distinguishable from the associated noise spectra.
The current noise is determined exactly and a rather sharp crossover
between flux noise in the MQT and charge noise in the CB regions is found
as the applied voltage is changed. Related results hold for the voltage noise
in current-biased junctions.

\end{abstract}

Generally, noise is considered undesirable and one searches for ways
to suppress it. However, occasionally the observation of noise may provide 
valuable information. The presence of shot noise in electrical transport 
indicates the discreteness of the charge carriers and the ratio between noise 
and current directly measures their charge. This fact was exploited to 
demonstrate the fractional charge in the fractional quantum Hall 
effect \cite{samin97,depic97}. 

Noise may also be helpful in identifying a transport mechanism. Tunnel
junctions often display a linear current-voltage characteristics and are
therefore indistinguishable from an ohmic resistor if only the current is
measured. On the other hand, noise measurements exhibit clear differences.
One finds shot noise in the first and Nyquist noise in the second case
corresponding to discrete and continuous charge transport, respectively.

An even more interesting situation arises, when different physical mechanisms
can occur as is the case for ultrasmall Josephson junctions. Such systems
have been proposed as building blocks for quantum computers \cite{makhl99}
and the operation of a superconducting box containing such a tunnel junction
as a qubit has been demonstrated \cite{nakam99}.

For a single ultrasmall Josephson junction at low temperatures, it has been
theoretically predicted that one may change from transport dominated by 
macroscopic quantum tunneling (MQT) to the regime of Coulomb blockade (CB) 
just by changing the applied voltage \cite{ingol99}. These two regimes are 
qualitatively different as in MQT the phase difference across the Josephson 
junction is a good quantum variable while CB is governed by the conjugate 
charge variable. We propose to study the noise properties in order to 
experimentally identify the transport mechanism.

\begin{figure} 
\onefigure[width=0.7\textwidth]{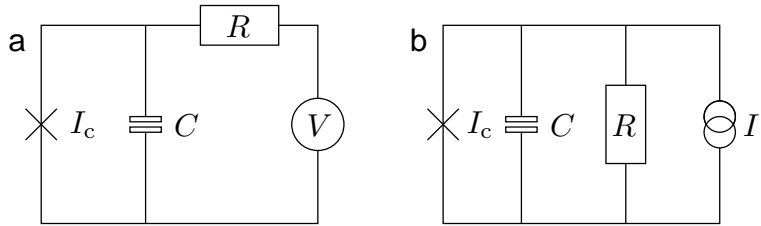} 
\caption{An ultrasmall Josephson junction is characterized by its critical 
current $I_{\rm c}$ and a capacitance $C$. a) Junction and ohmic resistor in
series are voltage-biased. b) Junction and ohmic resistor in parallel are
current-biased.}
\label{fig:circuit}
\end{figure}

We will discuss the noise properties of a voltage-biased as well as a 
current-biased small Josephson junction with the effective circuits shown
in figs.~\ref{fig:circuit}a and \ref{fig:circuit}b, respectively. The Josephson 
junction may be characterized by its critical current $I_{\rm c}$ and its 
capacitance $C$ which lead to two energy scales governing the behavior of 
the junction. The Josephson energy $E_{\rm J} = \hbar I_{\rm c}/2e$ determines 
the probability of Cooper pair tunneling while $E_{\rm c} =(2e)^2/2C$ is the 
charging energy of a capacitor carrying just one Cooper pair. The resistance
$R=\rho R_{\rm Q}$ of the external resistor may be taken relative to the 
resistance quantum $R_{\rm Q}=h/4e^2$. Typically, the resistance will be small,
i.e. $\rho\ll1$. In the following, we will be interested in the behavior of the 
junction at voltages of the order of $RI_{\rm c}$ much smaller
than the superconducting gap. Therefore, quasiparticle excitations may be
neglected.

Exact results for the current-voltage characteristics and current noise are
known \cite{fendl95,weiss96,weiss97} for some one-dimensional systems
with ohmic dissipation corresponding to an ideal ohmic resistor. However,
as can be seen from fig.~\ref{fig:circuit}, the external resistance is cut 
off by the junction's capacitance at high frequencies. For typical lead 
resistances, the cutoff frequency $1/RC$ is much larger than the frequency 
$(2e/\hbar)RI_{\rm c}$ corresponding to the typical voltages of interest and
the assumption of an ohmic resistor is sufficient for these voltages. This
implies an overdamped junction characterized by a McCumber parameter
$\beta_{\rm c}=(2e/\hbar)R^2 I_{\rm c} C\ll1$. In the following, we will 
focus on the overdamped regime.

We start with a discussion of the voltage-biased case (fig.~\ref{fig:circuit}a).
In the overdamped limit and $\rho<1$, the zero temperature current-voltage 
characteristics displays an almost linear rise of the current for small 
voltages reflecting the fact that nearly the entire applied voltage drops 
across the external resistor. There are, however, deviations due to macroscopic 
quantum tunneling which causes phase slips by quantum tunneling of the phase 
accoss the barrier of the Josephson potential. This is responsible for the 
voltage drop across the junction captured by a perturbation theory in 
$E_{\rm c}/E_{\rm J}$ yielding the current-voltage characteristics 
\cite{ingol99}
\begin{equation}
\langle I_{\rm J}\rangle = \frac{V}{R}\left[1-\sum_{n=1}^{\infty}c_n(1/\rho)
\left(\Gamma(1+\rho)^{1/\rho}\frac{{\rm e}^{\gamma}}{\pi^2\rho^2}
\frac{E_{\rm c}}{E_{\rm J}}\right)^{2n}
\left(\frac{V}{RI_{\rm c}}\right)^{2(1/\rho-1)n}\right]
\label{eq:ivmqt}
\end{equation}
where
\begin{equation}
c_n(\rho) = (-1)^{n-1}\frac{\Gamma(1+\rho n)\Gamma(3/2)}
{\Gamma(1+n)\Gamma(3/2+(\rho-1)n)}
\end{equation}
and $\gamma=0.577\dots$ is the Euler constant.

On the other hand, perturbation theory in $E_{\rm J}/E_{\rm c}$ yields the
current-voltage characteristics
\begin{equation}
\langle I_{\rm J}\rangle = \frac{V}{R}\sum_{n=1}^{\infty}c_n(\rho)
\left(\frac{\pi^2\rho^2{\rm e}^{-\gamma}}{\Gamma(1+\rho)^{1/\rho}}
\frac{E_{\rm J}}{E_{\rm c}}\right)^{2n\rho}
\left(\frac{V}{RI_{\rm c}}\right)^{-2(1-\rho)n}
\label{eq:ivcb}
\end{equation}
which describes incoherent tunneling of Cooper pairs across the oxide layer
of the Josephson junction. The leading-order behavior $I\sim V^{2\rho-1}$
is typical for Coulomb blockade which for $\rho>1$ manifests itself in a 
suppression of the current at low voltages. For $\rho<1$, this term would
correspond to a divergence at zero voltage and then (\ref{eq:ivcb}) can only
be valid for not too small voltages.

\begin{figure} 
\onefigure[width=0.6\textwidth]{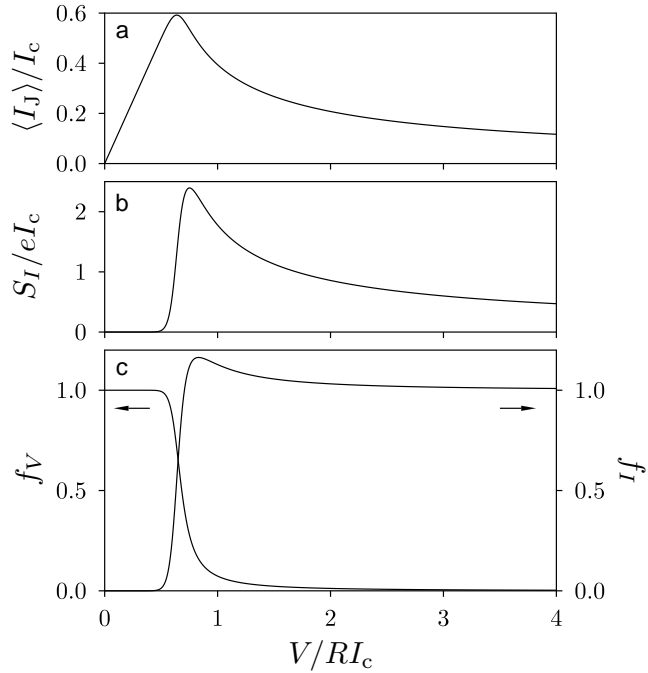} 
\caption{Current and current noise have been calculated for an ultrasmall 
Josephson junction with $E_{\rm c}=E_{\rm J}$ and an external resistance 
$R=0.1R_{\rm Q}$. a)  Current-voltage characteristics. b) Current noise as
a function of the voltage bias. c) Fano factors
for flux noise (left scale) and charge noise (right scale) appropriate in 
the MQT and CB regime, respectively.}
\label{fig:noise}
\end{figure}

In fact, the two series have a finite radius of convergence. For $\rho<1$, the 
expansions (\ref{eq:ivmqt}) and (\ref{eq:ivcb}) converge for low and high 
voltages, respectively. They join smoothly and provide a full description of 
a peaked current-voltage characteristics. To the left of the peak, transport
is therefore based on macrosopic quantum tunneling while to the right of the 
peak, we find the regime of Coulomb blockade. The two regimes, even though the 
underlying physics is very different, are related to each other by a duality 
transformation \cite{fendl95,weiss96,weiss97}.
A typical example for the current-voltage characteristics is depicted in
fig.~\ref{fig:noise}a for a Josephson junction with $E_{\rm c}=E_{\rm J}$ and 
a small environmental resistance $R=0.1R_{\rm Q}$. The peak at a voltage of 
order $RI_{\rm c}$ is a remnant of the dc Josephson effect of a classical 
Josephson junction.

The question now arises how to identify the two transport regimes without
making use of the theoretical results. We argue that a suitable way to 
achieve this goal is the observation of current noise 
\begin{equation}
S_I = \int_{-\infty}^{+\infty}{\rm d}t\langle 
\delta I_{\rm J}(t)\delta I_{\rm J}(0)
+\delta I_{\rm J}(0)\delta I_{\rm J}(t)\rangle\,.
\end{equation}
Here, $\delta I_{\rm J}$ denotes the deviation of the current $I_{\rm J}$ from 
its mean value $\langle I_{\rm J}\rangle$. 
The noise may be determined by following the same line of reasoning employed 
previously to calculate the noise in fractional quantum Hall bars 
\cite{fendl95,weiss96,weiss97}. The time evolution of the density matrix may
be written as a path integral on the Keldysh contour including an auxiliary
field coupling to the current operator. Arbitrary current expectation values
are then determined as functional derivatives of the path integral. Concrete
results like the series (\ref{eq:ivmqt}) and (\ref{eq:ivcb}) for the $I$-$V$ 
curves may be obtained in the so-called Coulomb gas representation 
\cite{weiss99} of the real-time path integral. Second order functional 
derivatives allow to determine the noise properties. 

In the limit of zero frequency, the results can be expressed in closed form and 
one obtains for the current noise
\begin{equation}
S_I=\frac{2eV}{1-\rho}(G-G_{\rm d}).
\label{eq:noise}
\end{equation}
Apart from the external voltage $V$ and the dimensionless resistance $\rho$,
the noise depends on the difference of absolute and differential conductance, 
$G=\langle I_{\rm J}\rangle/V$ and $G_{\rm d}=\partial\langle 
I_{\rm J}\rangle/\partial V$, where $\langle I_{\rm J}\rangle$ is the time 
averaged current. Note that in the case of 
an ideal supercurrent the external voltage drops entirely across the resistor. 
Then, the current-voltage characteristics is linear in the external voltage and 
the noise vanishes due to the fully coherent transport of Cooper pairs.
The result (\ref{eq:noise}) allows us to obtain the current noise in the
middle panel of fig.~\ref{fig:noise} from the current-voltage characteristics 
shown the upper panel. The current noise in fig.~\ref{fig:noise}c has been 
plotted as two different Fano factors. As will be explained in the following, 
these Fano factors are appropriate to identify the transport mechanisms.

The current noise both in the CB and MQT regimes may be 
understood in terms of Poissonian shot noise where transport of the 
appropriate quantity occurs at uncorrelated random times. The shot noise is 
given by the product of the transported quantity and the corresponding 
current.  In the CB regime, it is the charge flow of Cooper pairs which obeys 
Poissonian statistics. The current noise, $S_I=4e\langle I_{\rm J}\rangle$, 
is therefore proportional to the charge $2e$ of a Cooper pair and the average 
current $\langle I_{\rm J}\rangle$ through the Josephson junction. The Fano 
factor
\begin{equation}
f_I= \frac{S_I}{4e\langle I_{\rm J}\rangle}
\end{equation}
plotted in fig.~\ref{fig:noise}c clearly confirms the assumption of shot noise
since it is very close to one in the CB region.
On the other hand, in the regime of MQT, the charge flow becomes continuous and 
the corresponding shot noise is strongly suppressed. The change from the MQT
to the CB regime is indicated by a remarkably sharp rise of the Fano factor
$f_I$.

In the case of MQT, occasional phase slips lead to a voltage drop
$V_{\rm J} = (\hbar/2e)\dot\varphi$ across the junction and to voltage noise
\begin{equation}
S_V = \int_{-\infty}^{+\infty}{\rm d}t\langle \delta V_{\rm J}(t)
\delta V_{\rm J}(0) + \delta V_{\rm J}(0)\delta V_{\rm J}(t)\rangle\,.
\label{eq:sv}
\end{equation}
Since $V_{\rm J}=V-RI_{\rm J}$ and the external voltage does not fluctuate, the 
current noise is determined by (\ref{eq:sv}) via $S_I=S_V/R^2$. The assumption 
of Poissonian statistics of the phase slips then allows us to evaluate the 
current noise. During a phase slip the phase changes by $2\pi$ leading to an 
integrated voltage pulse $h/2e$. The current noise thus becomes
\begin{equation}
S_I=\frac{1}{R^2}\frac{h}{e}\langle V_{\rm J}\rangle\,.
\end{equation}
The corresponding Fano factor
\begin{equation}
f_V = \frac{eR^2}{h}\frac{S_I}{\langle V_{\rm J}\rangle}
\end{equation}
therefore allows to identify MQT as 
is shown in the left part of fig.~\ref{fig:noise}c. In contrast, in the CB 
regime the phase is strongly fluctuating and shot noise due to phase 
slips can no longer be detected. Again, the crossover between the two
regimes is very distinct.

The results discussed so far for the voltage-biased case may be rewritten for
a current-biased junction (fig.~\ref{fig:circuit}b). The voltage-biased case with applied 
voltage $V$ and current $I_{\rm J}$ through the junction can be transformed to 
the current-biased case with applied current $I$ and voltage drop $V_{\rm J}$
across the junction by means of the relations $I=V/R$ and 
$V_{\rm J}=V-RI_{\rm J}$. Then, the current noise (\ref{eq:noise}) turns into
voltage noise
\begin{equation}
S_V = \frac{2eRI}{1-\rho}\left(\frac{\partial\langle V_{\rm J}\rangle}{\partial
I}-\frac{\langle V_{\rm J}\rangle}{I}\right),
\label{eq:svcb}
\end{equation}
which depends on the difference of the differential and the absolute resistance
of the resistively shunted junction. 

\begin{figure} 
\onefigure[width=0.6\textwidth]{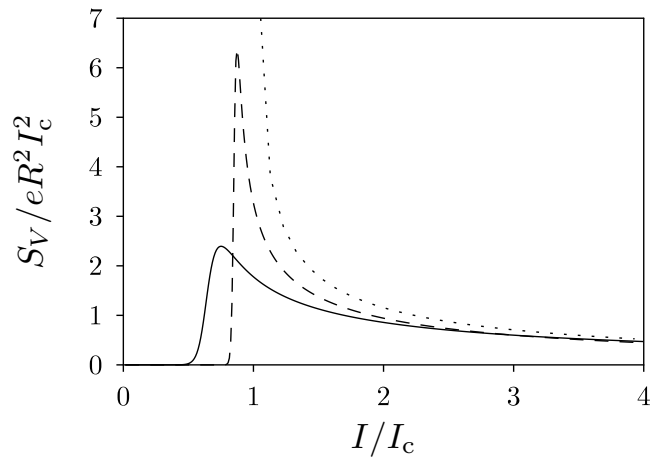} 
\caption{The voltage noise (\protect\ref{eq:svcb}) as a function of the bias
current is shown for a junction with $E_{\rm c}=E_{\rm J}$ and two external
resistances $\rho=0.1$ (full line) and $\rho=0.02$ (dashed line). The result
(\protect\ref{eq:sv0}) in the limit $\rho\to0$ is represented by the dotted
line.}
\label{fig:sv}
\end{figure}

It is instructive to make connection to results known in the limit $\rho\to 0$, 
where the current-voltage characteristics for $I>I_{\rm c}$ is given by
\cite{stewa68,mccum68}
\begin{equation}
V_{\rm J} = R\left(I^2-I_{\rm c}^2\right)^{1/2}.
\label{eq:vji0}
\end{equation}
From (\ref{eq:svcb}) one therefore finds for the voltage noise
\begin{equation}
S_V = 2eR^2I_{\rm c}^2\left(I^2-I_{\rm c}^2\right)^{-1/2}
\label{eq:sv0}
\end{equation}
in agreement with the results of Ref.~\cite{likha72}. While this result 
diverges when $I$ approaches the critical current $I_{\rm c}$, the expression 
(\ref{eq:svcb}) for $\rho>0$ yields a well-behaved voltage noise for the entire 
range of applied currents. Fig.~\ref{fig:sv} compares the voltage noise 
according to (\ref{eq:svcb}) for a junction with $E_{\rm c}=E_{\rm J}$ and 
finite shunt resistance $R$ corresponding to $\rho=0.1$ and $\rho=0.02$ with 
the result (\ref{eq:sv0}) for $\rho\to0$.
The divergence at $I=I_{\rm c}$ associated with the kink in the voltage-current
characteristics (\ref{eq:vji0}) is smoothed as the external resistance
increases.

In conclusion, we have studied noise properties of voltage-biased small 
junctions, which have been the subject of recent experimental investigations 
\cite{stein01}, as well as of the more standard current--biased junctions
employed in SQUID technology \cite{clark83}. Even though
we started from analytical results valid for the overdamped limit at zero
temperature, the reasoning leading to the Fano factors was completely
independent of these results. They were only needed to confirm the validity of 
the assumption of Poissonian statistics for charge transport and phase slips. 
One may therefore conclude that the observation of noise allows to determine 
the transport mechanism independently of a theoretical result for the 
current-voltage characteristics and the current noise. As a consequence, noise 
measurements may well be useful to identify the transport mechanism beyond the 
overdamped limit and the limit of zero temperature.

\acknowledgments
We would like to thank M.~H.~Devoret, D.~Esteve, H.~Saleur, C.~Urbina,
and U.~Weiss for numerous inspiring discussions and the Institute for 
Theoretical Physics at UCSB for hospitality during the workshop on 
``Nanoscience'' where this work was completed. This research was supported 
in part by the National Science Foundation under Grant No.\ PHY99-07949 and
the Deutsche Forschungsgemeinschaft through Sonderforschungsbereich 484.

\end{document}